\documentclass[prl,twocolumn,preprintnumbers,amsmath,amssymb,
floatfix,showpacs]{revtex4}
\usepackage{hyperref}
\usepackage{amsmath}
\usepackage[dvips]{epsfig}

\begin{document}
\title{Monte Carlo Simulations of the Unitary Bose Gas}
\author{Maurizio Rossi$^1$, Luca Salasnich$^1$, 
        Francesco Ancilotto$^{1,2}$ and Flavio Toigo$^{1,2}$} 
\affiliation{$^1$Dipartimento di Fisica e Astronomia "Galileo Galilei" and CNISM, 
                 Universit\`a di Padova, via Marzolo 8, 35122 Padova, Italy \\
             $^2$CNR-IOM Democritos, via Bonomea, 265 - 34136 Trieste, Italy} 

\begin{abstract} 
We investigate the zero-temperature properties of a diluted homogeneous Bose gas
made of $N$ particles interacting via a two-body square-well potential by performing 
Monte Carlo simulations.
We tune the interaction strength to achieve arbitrary positive values of the 
scattering length and compute by Monte Carlo quadrature the energy per particle $E/N$ 
and the condensate fraction $N_0/N$ of this system by using a Jastrow ansatz 
for the many-body wave function which avoids the formation of the self-bound 
ground-state and describes instead a (metastable) gaseous state with uniform density.
In the unitarity limit, where the scattering length diverges while the range of the 
inter-atomic potential is much smaller than the average distance between atoms, we find 
a finite energy per particle ($E/N=0.70\ \hbar^2(6\pi^2n)^{2/3}/2m$, with $n$ the number 
density) and a quite large condensate fraction ($N_0/N=0.83$).   
\end{abstract} 

\date{6.02.2014}
\pacs{03.75.Fi, 67.85.-d, 05.10.Ln}
\maketitle

One of the most intriguing topics in modern quantum physics is the characterization of 
the universal properties of an ultracold and dilute atomic gas in the so-called unitary 
regime \cite{bert}, i.e. when the two-body scattering length $a$ is tuned to very large 
values by using the Feshbach resonance technique \cite{fesh}, and the range $R$ of the 
inter-atomic potential is much smaller than the average distance $r_0$ between atoms 
\cite{braa}. 
It is now understood that the unitary regime is characterized by remarkably simple universal 
laws, arising from scale invariance, and has connections with fields as diverse as nuclear 
physics and string theory \cite{cast}. 
In the last years the unitary Fermi gas has been largely investigated both experimentally 
and theoretically \cite{book}, while its bosonic counterpart has been only marginally 
addressed theoretically \cite{peth,sal1,japs,lee2,sto1,sto2} because generally considered as 
experimentally inaccessible \cite{hoho}. 

Contrary to the case of Fermi gases, a Bose gas with attractive interactions is mechanically 
unstable at low $T$, and thus most of the studies have been focused only on repulsive Bose 
gases. 
However, in the strongly repulsive regime there is a huge increase of the three-body 
recombination rates close to a Feshbach resonance \cite{lee2,robe}, which makes very difficult 
to reach an equilibrium state. 
Very recent experimental observations \cite{liho,brem,corn,zora}, however, have put the seed 
for future investigations even for the degenerate Bose gas, showing that the three-body 
dynamics that spoils the unitary regime is slow enough with respect to the two-body one so that 
the degenerate Bose gas evolves dynamically on time scales fast compared to losses, thus allowing 
a unitary Bose gas to be experimentally created and probed dynamically. 
In spite of these promising results, however, the behavior of a Bose gas in this metastable 
regime is still not well understood and in the recent literature on the subject quite different 
predictions on its bulk properties are reported \cite{peth,japs,lee2,sto1,sto2}.

The bosonic unitary regime is a formidable challenge for many-body theories. 
Due to the strong interaction the standard mean-field theories are inadequate and the metastability 
of the system also rules out all those (ab-initio or not) microscopic theories that are explicitly 
devised to search for the ground state. 
In particular no attempt has been made yet to derive the equation of state of bosons at unitarity 
using microscopic quantum Monte Carlo approaches as done for fermion gases \cite{gio1,gio2}.
The reason is that a positive scattering length $a$ is associated with the presence of two-body 
bound states of energy $e_b=-\hbar^2/(2ma^2)$ in the interaction potential.  
This makes the gas-like state unstable and drives the system towards a self-bound ground state 
(cluster formation).

In this Letter we address the problem of the metastable unitary Bose gas by using quantum Monte Carlo 
method where the many-body wave function is based on a Jastrow ansatz which explicitly avoids the 
formation of the self-bound ground-state.
We compute the energy per particle $E/N$ and the condensate fraction $N_0/N$ of the metastable  
Bose gas by numerically simulating a large number of bosons interacting via a square-well two-body 
potential of radius $R$ in a periodically repeated cubic cell. 
We study the meta-stable state by tuning the value of the s-wave scattering length $a$ via the 
two-body potential parameters, keeping the system in the dilute regime $R/r_0 \ll  1$
(where $r_0 = (3/(4\pi n))^{1/3}$ is the average distance among particles and $n$ the number density).
In the weak-coupling regime ($a/r_0\ll  1$) we recover the familiar results for the weakly-interacting 
Bose gas \cite{land}. 
In the strong-coupling regime ($a/r_0 \gg 1$) we reach the unitarity limit finding a finite and positive 
energy per particle, $E/N=0.70\ \varepsilon_B$ ($\varepsilon_B=\frac{\hbar^2}{2m}(6\pi^2n)^{2/3}$ is the 
characteristic energy emerging at unitarity for a Bose gas) and a large condensate fraction $N_0/N=0.83$. 

Before giving the details of our calculations, we briefly review the approximate theoretical methods 
used so far to approach the problem of a Bose unitary gas, and quote their main results for the energy 
and condensate fraction.

One of the simplest method that provides a better insight than the mean-field approach without 
suffering of the limitation of full microscopic techniques is the lowest order constrained variational 
(LOCV) method \cite{locv}. 
The LOCV recipe is based on a Jastrow wave function where the pair function $f(r)$ for small distances 
is the exact solution $f_2(r)$ of the two--body Schr\"odinger equation, while it is set to $1$ beyond a 
certain healing length. 
In the unitary limit, LOCV method predicts for a condensed Bose gas a finite value for the energy per 
particle $E/N=\xi\varepsilon_B$ \cite{peth}, with $\xi=1.75$.
Other viable strategies which have been used to study this system are Renormalization Group (RG) 
\cite{lee2,sto2} and hypernetted chain (HNC) approximations \cite{sto1}, that have both a long history 
in the study of strongly correlated systems.
RG approach provides the values $\xi=0.39$ \cite{lee2} and  $\xi=0.85$ \cite{sto2}, while an extrapolation 
from intermediate to very large scattering lengths of results from an HNC approach gives $\xi\simeq0.67$ 
\cite{sto1}.
The value $\xi=0.48$ has also been proposed, based on a variational approach on the momentum distribution 
\cite{japs}.
The results for the condensate fraction $N_0/N$ are even more scattered: LOCV gives a null condensate 
fraction \cite{peth}, variational and RG arguments lead to $N_0/N\simeq0.5$ \cite{japs,sto2}, while HNC 
provides the value $N_0/N=0.75$ \cite{sto1}.  
The fact that the theoretical predictions are so scattered is a signature of the absence of a standard 
procedure to face the metastable nature of the Bose gas at unitarity.

In our Monte Carlo calculations we treat $N=500$ atomic Bosons in a cubic simulation box with periodic 
boundary conditions.
The system is governed by the Hamiltonian 
\begin{equation}
 {\hat H} = \sum_{i=1}^N -\frac{\hbar^2}{2m} \nabla_i^2 + 
\sum_{i<j=1}^N v(|{\vec r}_i-{\vec r}_j|)
\end{equation}
with the two-body potential $v(|{\vec r}_i-{\vec r}_j|)$ given by: 
\begin{equation}
 \label{square}
 v(r) = \left\{
               \begin{array}{ccc}
                 -U_0      & {\rm for} & 0<r<R   \\
                 0         & {\rm for} & r>R  
               \end{array}
        \right.
\end{equation}
where $R$ is the range of the potential and $-U_0=-\hbar^2k_0^2/m$ is the well depth. 
The corresponding scattering length reads $a = R [1 - \tan(k_0R)/(k_0R)]$ and the effective range is 
$r_e = R[1 - R^2/(3a^2) - 1/(k_0^2aR)]$ \cite{lipp}.
We have chosen $k_0$ in such a way that there is a single bound state in the potential well and that the 
scattering length is positive.
When computing the properties of the system in the unitary limit, which is the main goal of the present 
work, the inequality $r_e\ll  r_0\ll  a$ must be satisfied. 
In order to verify it we have considered $R/r_0$ values smaller than $0.01$ and $a/r_0$ as large as $10000$.
Notice that as $a$ diverges $r_e$ becomes equal to $R$.   

To construct the many-body wave function we rely on a standard Jastrow-Feenberg ansatz keeping explicit 
only the two body-correlations:
\begin{equation}
 \label{JWF}
 \psi_J(\vec r_1, \vec r_2, \dots, \vec r_N) = \prod_{i<j}f(|\vec r_i - \vec r_j|) \; . 
\end{equation}
Since  the temperature is zero in our simulations and the Bose gas is dilute, its physical properties are 
governed by two-body interactions (at least in the metastable unitary regime), and the low energy scattering 
of two particles can be safely approximated with the solution of the two body problem \cite{peth,sto1}. 
We thus construct the pair function $f(r)$ in Eq. (\ref{JWF}) starting from the exact solution of the two-body 
Schr\"odinger equation
\begin{equation}
 \label{2bSE}
 \left[-\frac{\hbar^2}{m}\frac{d^2}{dr^2} + v(r)\right]r f_2(r) 
= \varepsilon \ r f_2(r) \; , 
\end{equation}
where $\varepsilon = \hbar^2k^2/m$ is the smallest positive energy compatible with the imposed periodic 
boundary conditions.
For the potential \eqref{square}, the solution of \eqref{2bSE} reads:
\begin{equation}
 \label{f2squ}
 rf_2(r) = \left\{
                \begin{array}{ccc}
                  A\sin(\kappa r)      & {\rm for} & 0<r<R   \\
                  B\sin(k r + \delta)  & {\rm for} & r>R  
               \end{array}
        \right.
\end{equation}
where $\kappa^2 = k^2 + k_0^2$ and the parameters $A$, $B$ and $\delta$ are fixed by the matching conditions 
at $r=R$ and by normalization \cite{wfn1}.

In the limit where the range $R$ of the two-body potential goes to $0$ the interparticle interaction in 
Eq.~(\ref{2bSE}) can be replaced by the Bethe-Peierls boundary conditions \cite{bet1,peth} on the pair function
\begin{equation}
 \label{BPbc}
 \lim_{r\to 0}\frac{(rf_2(r))'}{rf_2(r)}= -\frac{1}{a}\ .
\end{equation}
In this case $f_2(r)$ is given by
\begin{equation}
 \label{f2pet}
 rf_2(r) = A\sin(k r + \delta) 
\end{equation}
where the parameter $\delta$ is now fixed by Eq.~\eqref{BPbc} and $A$ by the normalization \cite{wfn2}.

In order to account for many-body effects, which typically become relevant when $r$ is of the same order of 
$r_0$, $f(r)$ is smoothly joined with a constant at a certain distance $R_m$ \cite{peth}.
This is required also in order to account for the periodic boundary conditions imposed to the simulation box. 
With such a wave function however, when the scattering length diverges, the equilibrium configuration is not 
a uniform gas, as desired, but rather it is a compact cluster of atoms in equilibrium with the vacuum, 
whose radius is about 10%
of the effective range of the potential, and 0.001 times the
average distance $r_0$ in the uniform system.
This is due to a maximum in the probability density provided by $r^2f_2^2(r)$ at short distances (order of $R$) 
which favors configurations where the atoms are close to one another (dimers, trimers, etc.).
In order to keep the system in a uniform phase (metastable state) we should correct $f(r)$ at small distances 
to prevent particles from dwelling into this regions.
The underlyng idea originates from Feynman's comments on the construction of the ground state wave function for 
liquid $^4$He \cite{feyn}.

There are different ways to implement such a correction, going from imposing the three body repulsive condition 
in quantum Monte Carlo methods \cite{piat} to enforcing the equivalent hard sphere condition in the 
hyperradius formalism \cite{stec}. 
These methods need an explicit way to include higher order correlations, that are not known, thus requiring 
extra approximations. 
In our approach we simply set to zero the value of the pair function up to the outermost node $R_n$ of $f_2(r)$. 
This is indeed reasonable because, due to the extreme diluteness of the gas, in the metastable unitary 
regime the particle pairs should experience only the tails of the wave function. 
Our choice has also the advantage of keeping all the formalism in the two--body sector.

The pair function then reads 
\begin{equation}
\label{pair}
f(r) = \left\{
               \begin{array}{ccc}
                0      & {\rm for} & 0<r<R_n   \\
                f_2(r) & {\rm for} & R_n<r<R_m \\
                1      & {\rm for} & r>R_m\ .
        \end{array}
        \right.
\end{equation}
 
The continuity and boundary conditions on $f(r)$, plus the requirement that $f'(R_m) = 0$ fix all the free 
parameters in $f(r)$ except for $R_m$.
As already pointed out in Ref.~\cite{peth}, one cannot simply treat $R_m$ as a variational 
parameter, since this procedure would lead to the undesired minimizing value $R_m = 0$.
We thus choose to fix $R_m$ through a normalization condition $4\pi n\ \int_0^{R_m}r^2f_2(r)\ dr = 1$ as in 
the LOCV approach \cite{peth}. 
There are other possible choices for $R_m$, such as $R_m=L/2$ as used in Quantum Monte Carlo studies of 
unitary Fermi gases \cite{gio1,gio2}, but such a choice could lead in the present case to the unwanted 
condition $r_0<R_n$ for large values of $a$. 

Summarizing: we propose a many-body wave function of the Jastrow-Feenberg form Eq.(\ref{JWF}) with the two 
body correlation function obeying three basic requirements, i.e. (i) it provides long range correlations as dictated 
by an attractive short range potential with the actual scattering length $a$; (ii) it keeps the density
uniform by preventing the formation of clusters; (iii) it is normalized while keeping the position of the last 
node of the actual two-body scattering wave function.

\begin{figure}
 \epsfig{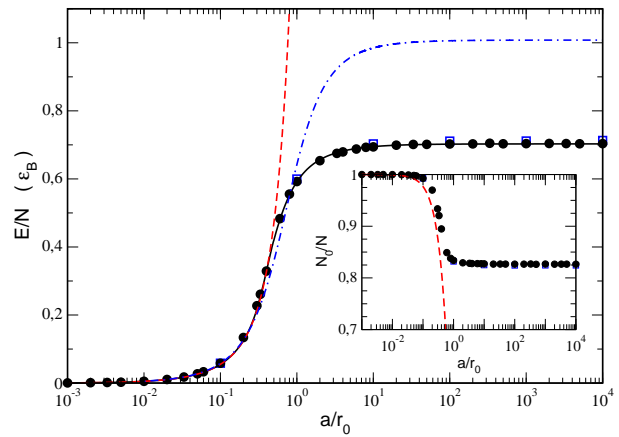}
 \caption{(Colors online) Energy per particle $E/N$ (in units of $\varepsilon_B$) and
          condensate fraction $N_0/N$ (inset) as a function of the scaled scattering 
          length $a/r_0$ for two values of the two body potential range $R/r_0$: 
          $0.01$ (open squares) and $0$ (filled circles). 
          The dashed line is the mean-field prediction of Bogoliubov with LHY correction 
          \cite{lhyc}, and the dot dashed line is the fit of Ref. \cite{sal2} to QMC 
          results for the unitary Fermi gas \cite{gio1}.
          The solid line is the fit of our Monte Carlo data as described in the text.}
 \label{fig1}
\end{figure}

In Fig.~\ref{fig1} we report the calculated energy per particle 
\begin{equation}
\varepsilon = \frac{E}{N} = 
\frac{1}{N}\frac{\langle \psi_J|{\hat H}|\psi_J\rangle}
{\langle \psi_J|\psi_J\rangle} 
\end{equation}
as a function of the scaled scattering length $a/r_0$ for two different values of the scaled two-body 
potential range $R/r_0$.
The MC simulations used to compute $\varepsilon$ never count less than $2\times10^6$ sampled configurations, 
and both the sparse and the block averaging techniques \cite{kalo} have been adopted to prevent 
correlations among the sampled configurations. 

The open squares are the results obtained at $R/r_0=0.01$ with $f_2(r)$ given by Eq.~\eqref{f2squ}, while 
the filled circles are obtained with $R/r_0=0$, i.e. with $f_2(r)$ given by Eq.~\eqref{f2pet}. 
The two sets of data are very close, showing that with $R/r_0=0.01$ the system is indeed dilute and displays 
(universal) properties which depend only on the s-wave scattering length $a$ \cite{braa,book}. 

We have also considered an alternative form for $f(r)$, which instead of being strictly zero in the range 
$0<r<R_n$ is smoothly connected to a third order polynomial in $r=R_n$ and fulfills the condition $f(0)=0$.
This leaves a free parameter that can be variationally optimized.
We found however that the resulting energy is always larger than the one obtained with \eqref{pair}.
In particular, the variational optimization returns a wave function as flat (and as small)
as possible within the range $0<r<R_n$, thus confirming the quality of our ansatz \eqref{pair}.

In the weakly interacting regime ($a/r_0\ll  1$) our results, as shown in Fig.~\ref{fig1}, agree with the 
well known universal Bogoliubov prediction \cite{bogo} with the Lee, Huang and Yang (LHY) correction \cite{lhyc}:
$\varepsilon_{\rm LHY}(x)=\varepsilon_B\left(\frac{4}{3\pi^2}\right)^{1/3}x\left[1+\frac{128}{15\sqrt{\pi}}
\sqrt{\frac{3}{4\pi}}x^{3/2}\right]$, where $x=a/r_0$. 
In the strong-coupling regime ($a/r_0\gg 1$) our data reach a plateau in a way that is qualitatively 
similar to the behavior founded for a unitary Fermi gas on the BEC side of BCS-BEC crossover (also shown for
comparison in Fig.~\ref{fig1}), but the 
convergence is to a lower value, namely $\varepsilon/\varepsilon_B = 0.70$.
This value is well below the LOCV prediction 1.75 based on the LOCV method \cite{peth}, and is slightly 
larger than the average value of RG approaches \cite{lee2,sto2}, the HNC result \cite{sto1} and the 
variational estimate of \cite{japs}. 

The obtained MC data are well interpolated by the function
\begin{equation}
 \label{fit}
 \varepsilon(x)/\varepsilon_B = \left\{
                                \begin{array}{ccc}
                                 \varepsilon_{\rm LHY}(x)/\varepsilon_B + ax^3        & {\rm for} & x<0.3     \\
                                 c_3x^3 + c_2x^2 + c_1x + c_0                         & {\rm for} & 0.3<x<0.5 \\
                                 b_0 + b_1\tanh\left(b_2/x - 1 \right)                & {\rm for} & x>0.5
                                \end{array}
                                \right.
\end{equation}
and a fit procedure provides the values $a = 0.21$, $b_0=0.45$, $b_1= -0.33$ and $b_2=0.54$, while
the parameters $c_3=-6.64$, $c_2=7.16$, $c_1=-1.46$ and $c_0=0.19$ are fixed by smoothness and continuity 
constraints in $r=0.3$ and 0.5 .
$\varepsilon(x)$ is shown with a solid line in Fig.~\ref{fig1}.

\begin{figure}
 \epsfig{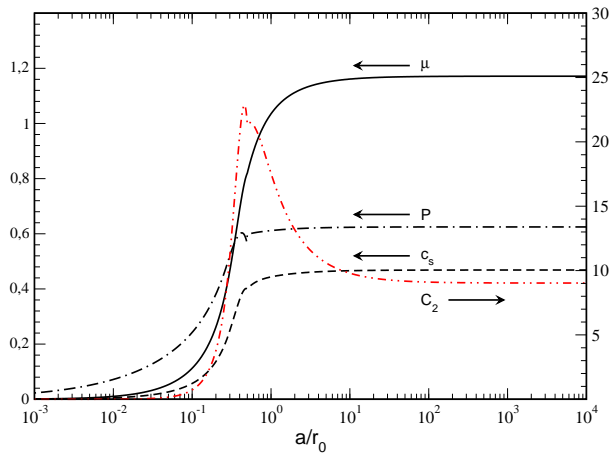}
 \caption{(Colors online) Chemical potential $\mu$ (in units of $\varepsilon_B$), 
          pressure $P$ (in units of  $n\varepsilon_B$), sound velocity $c_s$ (in 
          units of $v_B=\sqrt{2\varepsilon_B/m}$) and two body contact density $C_2$ (in 
          units of $n^{4/3}$) as a function of the scaled scattering length $a/r_0$ 
          as obtained from the parametrization \eqref{fit} of the equation of state.
          Arrows indicate the axis with the corresponding scale.}
\label{figN}
\end{figure}

The parametrization \eqref{fit} of the equation of state allows to obtain other useful quantities
via standard thermodynamical relations, as for example the chemical potential 
$\mu = \partial_n(n\varepsilon )$, the pressure 
$P = n^2\partial_n\varepsilon$, the sound velocity,
$c_s^2 = n/m\ \partial_n\mu$, and also the Tan's two-body contact density
\cite{tans} $C_2 = (8\pi n ma^2/\hbar^2) {d\varepsilon /da}$ 
which describes the $1/k^4$ tail of the momentum distribution $\rho(k)$ at large momenta. 
Our results are reported in Fig.~\ref{figN}.
Note that at unitarity $C_2=\alpha n^{4/3}$, with $\alpha = 9.02$.
This value compares acceptably well with previous theoretical estimates 
$\alpha=10.3$ \cite{sto1}, 32 \cite{sto2} and 12 \cite{skye}, and is a factor of 2
smaller than the value extrapolated from experimental results on a trapped gas in local 
density approximation, $\alpha=22$ \cite{tans}.
 
The many-body wave function \eqref{JWF} gives direct access also to the one-body density 
matrix $\rho_1(|\vec r - \vec r'|)$ whose limiting value at large distances
provides the condensate fraction $N_0/N$ \cite{lipp}.
In the inset of Fig.~\ref{fig1} we plot the behavior of $N_0/N$ as a function of $a/r_0$ 
for two values of $R/r_0$.
Our data follow the Bogoliubov prediction 
$\frac{N_0}{N} = 1 - \frac{8}{3\sqrt{\pi}}\left(\frac{3}{4\pi} x^3\right)^{1/2}$
(dashed line in Fig.~\ref{fig1}) up to about $a/r_0 = 0.01$,
while in the unitary limit converge to a constant value $N_0/N=0.83$.
The LOCV method predicts in the same limit $N_0/N=0$ \cite{peth} while the 
HNC value, $N_0/N=0.78$ \cite{sto1} is compatible with our result.
For completeness, we remind that the RG method developed in Ref.~\cite{sto2} and the
variational approach in Ref.~\cite{japs} suggest instead $N_0/N\sim 0.5$. 

In conclusion, we have studied the zero temperature unitary Bose gas via a Jastrow ansatz 
on the many-body wave function which avoids the formation of the self-bound ground-state
and then computed the energy per particle and the condensate fraction by Monte Carlo 
quadrature. 
In the unitary limit we have found a finite value both for the energy per particle and 
for the condensate fraction. 
This is a clear signature of a universal behavior in which the properties of the system
depend only on the average distance among the particles encoded in their density.
The fact that the universal value of the energy per particle for the unitary gas is lower
for Bosons than for Fermions is not completely unexpected, since the antisymmetry of the
Fermionic wave function results in an effective excluded volume effect that increases the
energy \cite{mull}.  
From the Monte Carlo data of the energy per particle we have also derived the chemical 
potential, the pressure, the sound velocity, and the contact density as a function of the 
s-wave scattering length. 
We believe our predictions can be tested with the ongoing experiments \cite{brem,corn,zora} 
on ultracold vapors of bosonic alkali-metal atoms.  

The authors thank D.E. Galli and G. Bertaina for useful discussions and suggestions. 
The authors acknowledge for partial support Universit\`a di Padova (grant No. CPDA118083), 
Cariparo Foundation (Eccellenza grant 11/12), and MIUR (PRIN grant No. 2010LLKJBX).

\end{document}